\documentstyle[epsfig]{elsart}              

\begin{document}
\begin{frontmatter}
%%%%%%%%%%%%%%%%%%%%%%%%%%%%%%%%%%%%%%%%%%%%%%%%%%%%%%%%%%%%%%%%%%%%%%%%        
                                                            
\title{Short-time dynamics of the positional order   
of the  two-dimensional hard disk system} 
 
\author{A. Jaster}
\address{Universit\"{a}t - GH Siegen, D-57068 Siegen, Germany}      
\date{\today}
                  
%%%%%%%%%%%%%%%%%%%%%%%%%%%%%%%%%%%%%%%%%%%%%%%%%%%%%%%%%%%%%%%%%%%%%%%%        

\maketitle
\begin{abstract}

We investigate the  positional order 
of the two-dimensional hard disk model with short-time dynamics
and equilibrium simulations. The melting density and
the critical exponents $z$ and $\eta$  
are determined. Our results rule out a phase transition as predicted by the
Kosterlitz-Thouless-Halperin-Nelson-Young theory as well as
a first-order transition.
\end{abstract}

\keyword
Solid-liquid transition; Hard disk model; Short-time dynamics; 
Non-equilibrium kinetics
\endkeyword

\end{frontmatter}
%%%%%%%%%%%%%%%%%%%%%%%%%%%%%%%%%%%%%%%%%%%%%%%%%%%%%%%%%%%%%%%%%%%%%%%%        
%%%%%%%%%%%%%%%%%%%%%%%%%%%%%%%%%%%%%%%%%%%%%%%%%%%%%%%%%%%%%%%%%%%%%%%%  
Usually, numerical measurements of critical exponents are carried
out from simulations in equilibrium using Monte Carlo (MC)
techniques. Such simulations at or near the phase transition point,
except for some special cases \cite{WOLF}, are affected by the
critical slowing down. Recently, Janssen, Schaub and Schmittmann \cite{JASCSC}
proposed an alternative, which allows their determination from the 
short-time dynamics. They discovered that a system with non-con\-served order
parameter (model A \cite{HOHHAL}) quenched from a high temperature 
state to the critical temperature  shows universal short-time behaviour.
This  sets in after a microscopic time scale $t_{\mathrm{mic}}$,
during which the non-universal behaviour is swept away.
Starting from a small value of the order parameter $m_0$ the order increases 
with a power law 
$M(t) \sim m_0 \, t^\theta$,
where $\theta$ is a new dynamic exponent. This short-time behaviour was
supported by a number of MC simulations \cite{HUSE,ZHENG}. 
These investigations also provide
a possibility to determine the conventional critical exponents. The
exponents can be calculated from the time evolution of the
second moment of the order parameter, the cumulant and
additional observables, which evolve also with a power law.
Since the simulations are performed in the early part of the dynamics,
this method may  eliminate critical slowing down.

While universal short-time behaviour was first seen when starting from
unordered states, short-time dynamical scaling can also  be found 
starting from the ordered state ($M(t=0)=1$). In the latter case,  
no analytical  calculations exist, but MC simulations have been performed
\cite{STAUFFER,LISCZH,ZHENG}. 
Again, one can use the  short-time
behaviour to calculate the critical exponents, except for the new exponent
$\theta$. In this letter, we use the dynamic relaxation of the
two-dimensional hard disk model starting from the ordered state to 
calculate the critical exponents
for the positional order.
Also this is the first time that the dynamic evolution in the early time is
studied for a non-lattice model.

The nature of the two-dimensional melting transition is a long unsolved
problem \cite{STRAND,GLACLA}. The 
Kosterlitz-Thouless-Halperin-Nelson-Young (KTHNY) theory \cite{KTHNY}
predicts two continuous transitions. The first transition occurs
at  temperature $T_{\mathrm{m}}$
when the solid (quasi-long-range positional order, long-range
orientational order) undergoes a dislocation unbinding transition
to the hexatic phase (short-range positional order, quasi-long-range
orientational order). The second transition is the 
disclination unbinding transition 
(at $T_{\mathrm{i}}$) which transforms this 
hexatic phase into an isotropic phase (short-range positional and 
orientational order). There are several other theoretical
approaches to the transition. An alternative scenario
has been proposed by Chui \cite{CHUI}. He presented a theory via 
spontaneous generation of grain boundaries, i.e.\ collective excitations 
of dislocations, and predicted a conventional 
first-order phase transition from the solid to the isotropic
phase. In this case,  a region exists where both phases coexist
instead of a hexatic phase.
Even for the simple hard disk system no consensus about the nature of the
transition  has been established. 
 
A number of simulations of the two-dimensional 
hard disk model in equilibrium have been performed.
The melting transition of the hard disk system was 
first seen in a computer simulation by Alder and Wainwright 
\cite{ALDWAI}. They used a system of 870 disks and molecular
dynamics methods (constant number of particles $N$, 
volume $V$ and energy $E$ simulations)
and found that this system undergoes a first-order phase
transition. But the results of such small systems are
affected by large finite-size effects. Recent simulations used 
MC techniques either with constant volume 
($NVT$ ensemble) \cite{ZOLCHE,WEMABI,MIWEMA,JASTER} or constant
pressure ($NpT$ ensemble) \cite{LEESTR,FEALST}. 
The analysis of Zollweg and Chester \cite{ZOLCHE} for the pressure gave
an upper limit for a first-order phase transition, but is
compatible with all  scenarios. 
Lee and Strandburg \cite{LEESTR} 
used isobaric MC simulations and 
found a double-peaked structure in the volume distribution.
Lee-Kosterlitz scaling led them to conclude that the phase transition
is of first order. However, the data are not in the scaling region,
since their largest system contained only 400 particles.
MC investigations of the bound orientational order
parameter via finite-size scaling with the block analysis technique 
of 16384 particle systems were done by
Weber, Marx and Binder \cite{WEMABI}. They also  
favoured a first-order phase transition. In contrast to this, 
Fern\'{a}ndez, Alonso and Stankiewicz \cite{FEALST}\footnote{For
a critical discussion of this work see Ref.\ \cite{WMFAS}.} 
predicted a one-stage continuous melting
transition, i.e.\ a scenario with a single continuous transition.
Their conclusions were based on the examination of the bond orientational 
order parameter in very long runs of different systems up to 15876
particles and hard-crystalline wall boundary conditions.
Mitus, Weber and Marx \cite{MIWEMA} studied 
the local structure of a system with $4096$ hard disks.
From the linear behaviour of a local order parameter they derived
bounds for a possible coexistence region.
Finally, we showed  by MC simulations in the
$NVT$ ensemble \cite{JASTER} that  a one-stage transition 
can be ruled out. The results of the bond orientational order parameter 
were compatible with the KTHNY predictions or a weak first-order transition.

In this letter, we present results for the positional order of the hard
disk model obtained through MC simulations
near the melting transition $\rho_{\mathrm{m}}$
and in the solid phase ($\rho > \rho_{\mathrm{m}}$). 
First we use traditional simulations
in equilibrium ($NVT$ ensemble)
to locate the melting density $\rho_{\mathrm{m}}$ 
and to calculate the critical exponent $\eta$. After that, we investigate
the dynamic relaxation at and above $\rho_{\mathrm{m}}$ (model C) 
and calculate the critical exponents $z$ and $\eta$
from the short-time behaviour.
We use always periodic boundary conditions and 
a rectangular box with ratio $\sqrt{3}:2$,
which is necessary since we start the relaxation from the
ordered state. The disk diameter is set
equal to one so that the lengths of the system are given by 
$L$ and $2 L/\sqrt{3}$, where $L=\sqrt{\sqrt{3} N/ 2 \rho}$. 
For the simulations of the dynamic relaxation we 
use the conventional (local) Metropolis algorithm and
choose the new positions of the particles (for symmetry reasons)
with equal probability within a circle centered about
its original position.  
Simulations in  equilibrium are performed with a non-local 
Metropolis updating algorithm \cite{JASTER1}.
In this case, the new positions of the particles are chosen as usual
within a square.

The $k$th moment of the 
positional order parameter $\psi_{\mathrm{pos}}$ is defined as
\begin{equation}
{\psi_{\mathrm{pos}}}^k = \left \langle \left | 
\frac{1}{N} \sum_{i=1}^N \exp ( \mathrm{i} \, \vec{G} \cdot \vec{r}_i )
\right | ^k \right \rangle \ ,
\end{equation}
where $\vec{G}$ is a reciprocal lattice vector and $\vec{r}_i$ 
denotes the position of  particle $i$. The magnitude of $\vec{G}$ 
is given by $2 \pi /a$, where $a=\sqrt{2/\sqrt{3}\rho}$ is the average 
lattice spacing. The orientation of
$\vec{G}$ was defined in two different ways, 
which lead to two different $\psi_{\mathrm{pos}}$. In a first 
definition we fix the direction of $\vec{G}$ to that of  
a reciprocal lattice vector of the perfect crystal (which are unique due
to the boundary condition of a rectangular box of ratio $2:\sqrt{3}$).
The reason is that large crystal tilting (with the given
boundary conditions) is not possible 
if we are in or near the solid phase, while
small fluctuations of $\vec{G}$ are ignored.
However, we use a second definition, 
where we determine the orientation of the
crystal from the global bond orientation
\begin{equation}
\label{eq_psi6}
\psi_6 =  \frac{1}{N} \sum_{i=1}^{N} 
\frac{1}{N_i} \sum_{j=1}^{N_i} \exp ( 6 \, \mathrm{i}\, \theta_{ij} )
\ .
\end{equation} 
The sum on $j$ is over the $N_i$ neighbours of the 
particle $i$ and $\theta_{ij}$ is the angle between the bond
formed by particles $i$
and $j$ and an arbitrary but fixed reference axis. 
Neighbours are obtained by the Voronoi construction. 
Since we try to determine the global orientation of a single 
configuration, no average is taken in Eq.\ (\ref{eq_psi6}).
With $\psi_6 \sim \exp(6 \,\mathrm{i}\, \alpha)$,
we define the angle $\alpha$ ($0\le \alpha < \pi/3 $) and therefore
the orientation of the crystal and $\vec{G}$. 
Of course, the definition of $\vec{G}$ is only valid in a strict manner
if the bond orientational correlation length $\xi_6$ is infinite
(i.e.\ if we are in the hexatic or solid phase).
Simulations show that this is probably the case \cite{JASTER}.
However, if $\xi_6$ is finite, we have at least the case that 
$\xi_6 \gg L$, since we simulate near the melting transition. Thus we have 
also  in this case some kind of global orientation\footnote{Recent 
results concerning the
liquid-solid structure of two-dimensional liquids can be found in Ref.\
\cite{PAMIRA}.}.
The simulations showed that the difference between the two definitions 
is negligible in the solid phase, while it is getting important below the
transition point. 

\begin{figure}
\begin{center}
\mbox{\epsfxsize=9.0cm
\epsfbox{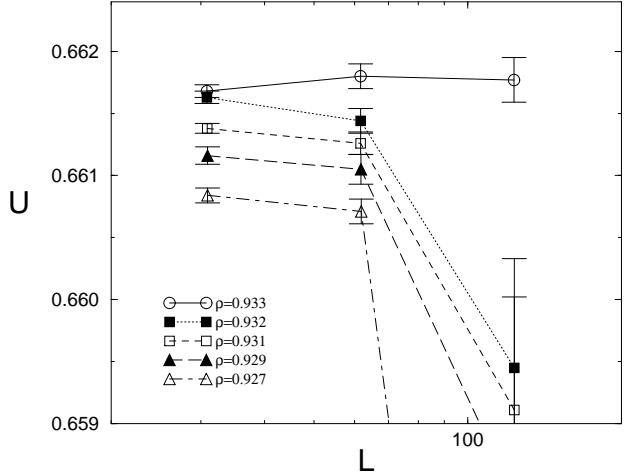}}
\end{center}
\caption{\label{fig_pos_FSS_U}
FSS of the cumulant in the vicinity of 
$\rho_{\mathrm{m}}$.}
\end{figure}

The positional correlation is quasi-long-ranged in the whole solid phase.
Therefore, the fourth-order cumulant
\begin{equation}
U=1-\frac{{\psi_{\mathrm{pos}}}^4}{3 
\left ( {\psi_{\mathrm{pos}}}^2 \right ) ^2}
\end{equation}
(measured in the equilibrium) is independent of the size of the system 
for $\rho \ge \rho_{\mathrm{m}}$. We use this finite-size scaling (FSS) 
behaviour to locate the melting density
$\rho_{\mathrm{m}}$ with simulations in equilibrium.
Our results for the dependence of $U$ on $L$ for $N=32^2$, $64^2$
and $128^2$ are shown in Fig.~\ref{fig_pos_FSS_U}. 
We use the second definition of $\psi_{\mathrm{pos}}$ (varying  orientation
of $\vec{G}$),
since most of the simulations are performed below $\rho_{\mathrm{m}}$.
However, the other definition leads to comparable results.
The figure shows that the scale invariance of $U$ yields a 
melting density of $\rho_{\mathrm{m}} \approx 0.933$ (in reduced units). 
Since there is a tendency of the slope to decrease with larger systems,  
scale invariance may actually take place at a slightly larger density.
However, even the value $\rho_{\mathrm{m}} =0.933$ is larger than the
upper limit of a possible liquid-solid tie line of 
$\rho_{\mathrm{s}} =0.912$ \cite{MIWEMA} and 
$\rho_{\mathrm{s}} =0.904$ \cite{ZOLCHE}, respectively.
The first result \cite{MIWEMA} was obtained by examining the behaviour of
a local bond orientational order parameter as a function of the density.
The second limit \cite{ZOLCHE} was determined from a study of the 
pressure as a function of the density. Both simulations do not rule out
a (one-stage or two-stage) continuous transition, but give limits for a
conventional first-order transition from the liquid to
the solid phase. However, for such a first-order
scenario the melting
density $\rho_{\mathrm{m}}$ (as determined from FSS of the
positional order) must lie
in the coexistence phase, i.e.\ below $\rho_{\mathrm{s}}$. 
Therefore, we can rule out a single first-order transition. 

%%%%%%%%%%%%%%%%%%%%%%%%%%%%%%%%%%%%%%%%%%%%%%%%%%%%%%%%%%%%%%%%%%%%%%%% 
%%%%%%%%%%%%%%%%%%%%%%%%%%%%%%%%%%%%%%%%%%%%%%%%%%%%%%%%%%%%%%%%%%%%%%%%
From the scaling relation
${\psi_{\mathrm{pos}}}^2 \sim L^{-\eta}$
we extract the critical exponent $\eta$ by a fit of all three
system sizes. At $\rho=0.933$ we get
$\eta=0.200(2)$, where the three points lie within statistical errors on 
a straight line. This  result is incompatible with
the prediction of the KTHNY theory of
$1/4 \le \eta(\rho_{\mathrm{m}}) \le 1/3$ \cite{KTHNY}.
Thus we can rule out the KTHNY scenario, because of the measured  
exponent $\eta$. If our $\rho_{\mathrm{m}}$ would be determined too
low, this result remains  unchanged.
We performed also  simulations
in equilibrium  at $\rho=1.0$ (solid phase) to verify
the scale invariance of $U$ above $\rho_{\mathrm{m}}$.
FSS yields $\eta=0.0791(6)$.

We now come to the dynamic relaxation of the hard disk system
starting from the ordered state. Simulations of the short-time dynamics 
of systems with quasi-long-range order were performed for the
6-state clock model \cite{CZERIT}, the XY model \cite{OKSCYAZH,LUOZHE}
and the fully frustrated XY model \cite{LUSCZH,LUOZHE}.
From the scaling form of the second moment of the order parameter at or above 
$\rho_{\mathrm{m}}$ 
\begin{equation}
{\psi_{\mathrm{pos}}}^2(t,L) = b^{-\eta}  
{\psi_{\mathrm{pos}}}^2(b^{-z}t,b^{-1}L) 
\end{equation}
one obtains that the
time dependence for sufficient large $L$ 
is given by a power law of the form \cite{ZHENG}
\begin{equation}
\label{EqPLpsi2}
{\psi_{\mathrm{pos}}}^2(t) \sim t^{-\eta/z} \ .
\end{equation}
Accordingly, FSS analysis of the
time-dependent cumulant 
\begin{equation}
\tilde{U}(t)= \frac{{\psi_{\mathrm{pos}}}^4(t)}{
\left ( {\psi_{\mathrm{pos}}}^2(t) \right ) ^2} -1
\end{equation}
leads to
\begin{equation}
\tilde{U}(t) \sim t^{d/z} \ .
\end{equation}
Therefore, one can determine the dynamic critical exponent $z$
from $\tilde{U}(t)$ and then, with $z$ in hand, the static exponent
$\eta$ from the behaviour of ${\psi_{\mathrm{pos}}}^2(t)$.
In principle, one can also try to determine $\rho_{\mathrm{m}}$
from the short-time dynamic,
as it was done for a second order transition \cite{SCHZHE}.
However, this determination  is difficult 
in systems with a KT-like transition  \cite{OKSCYAZH,ZHENG}.

In the following we perform MC simulations at $\rho_{\mathrm{m}}$
and in the solid phase ($\rho=1.0$) to investigate the time evolution
of ${\psi_{\mathrm{pos}}}^2$ and $\tilde{U}$. For short-time simulations
we use the first definition of  $\vec{G}$ (fixed direction), since large
crystal tilting is not possible if we study only the early part of
the evolution. However, we checked that both definitions lead to
similar results. Starting from the ordered
state ($\psi_{\mathrm{pos}}=1$), i.e.\ with a perfect crystal with lattice
spacing $a$, we release the system to evolve with the MC dynamic according
to the Metropolis algorithm. We use systems of $8^2$, $16^2$, $32^2$,
$64^2$ and $128^2$ hard disks and measure the observables up to 5000 
MC sweeps. The average is taken over ${\cal O}(5000)$ samples for 
$N=128^2$ and ${\cal O}(300\,000)$ samples for $N=8^2$.
The critical exponents are determined by least squares fits with a 
power law ansatz in the time interval $t=(t_{\mathrm{min}},t_{\mathrm{max}})$. 
Statistical errors are calculated by
dividing the data into different subsamples. Systematic errors are
estimated by the results of different system sizes and different time
intervals, i.e.\ we examined the dependency of the slope from
the fitted interval $t=(t_{\mathrm{min}},t_{\mathrm{max}})$
and number of particles $N$. The quoted error is a sum of the
statistical and systematic error.
%%%%%%%%%%%%%%%%%%%%%%%%%%%%%%%%%%%%%%%%%%%%%%%%%%%%%%%%%%%%%%%%%%%%%%%% 
\begin{figure}[t]
\begin{center}
\mbox{\epsfxsize=6.5cm
\epsfbox{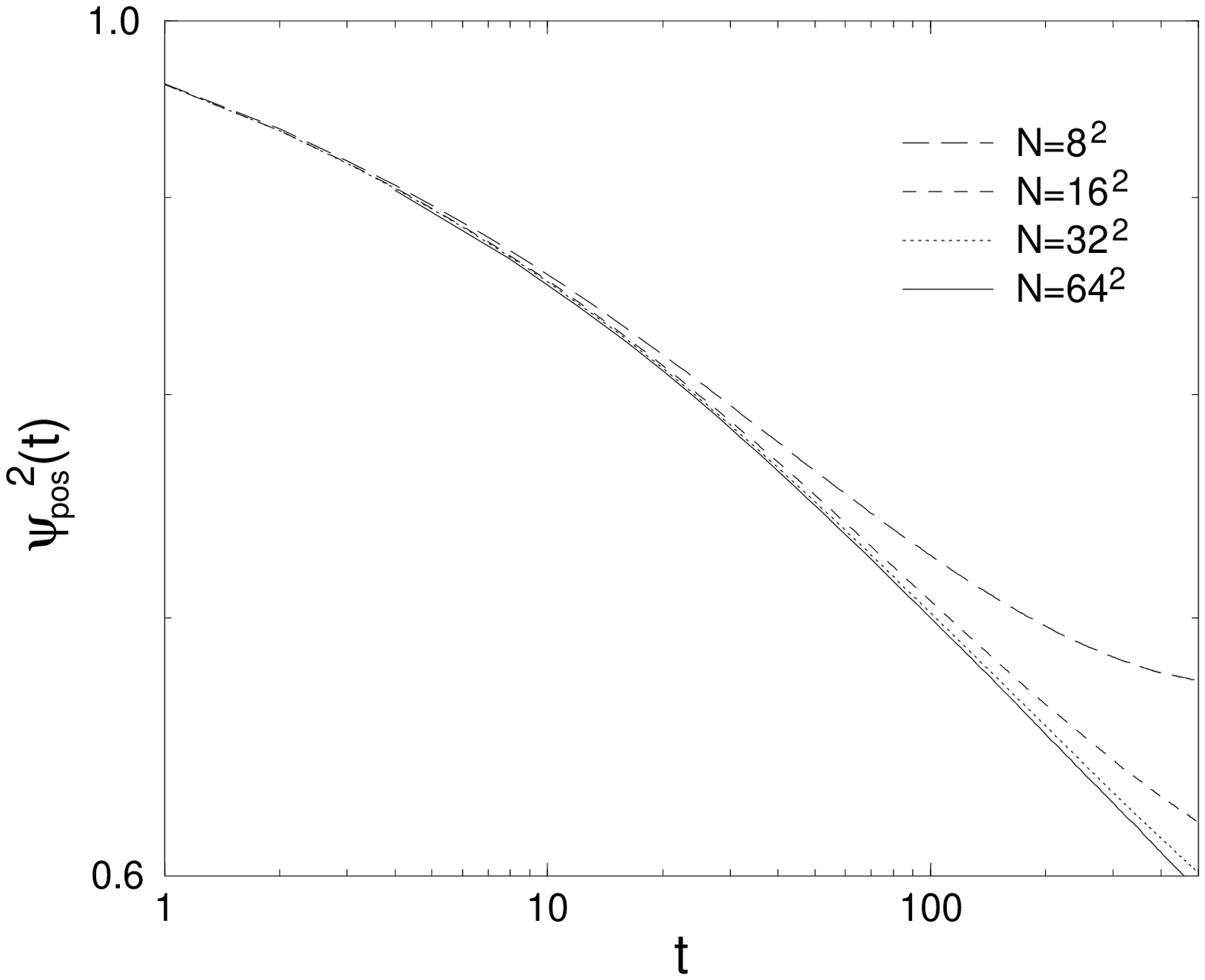}}
\hspace{1.0eM}
\mbox{\epsfxsize=6.5cm
\epsfbox{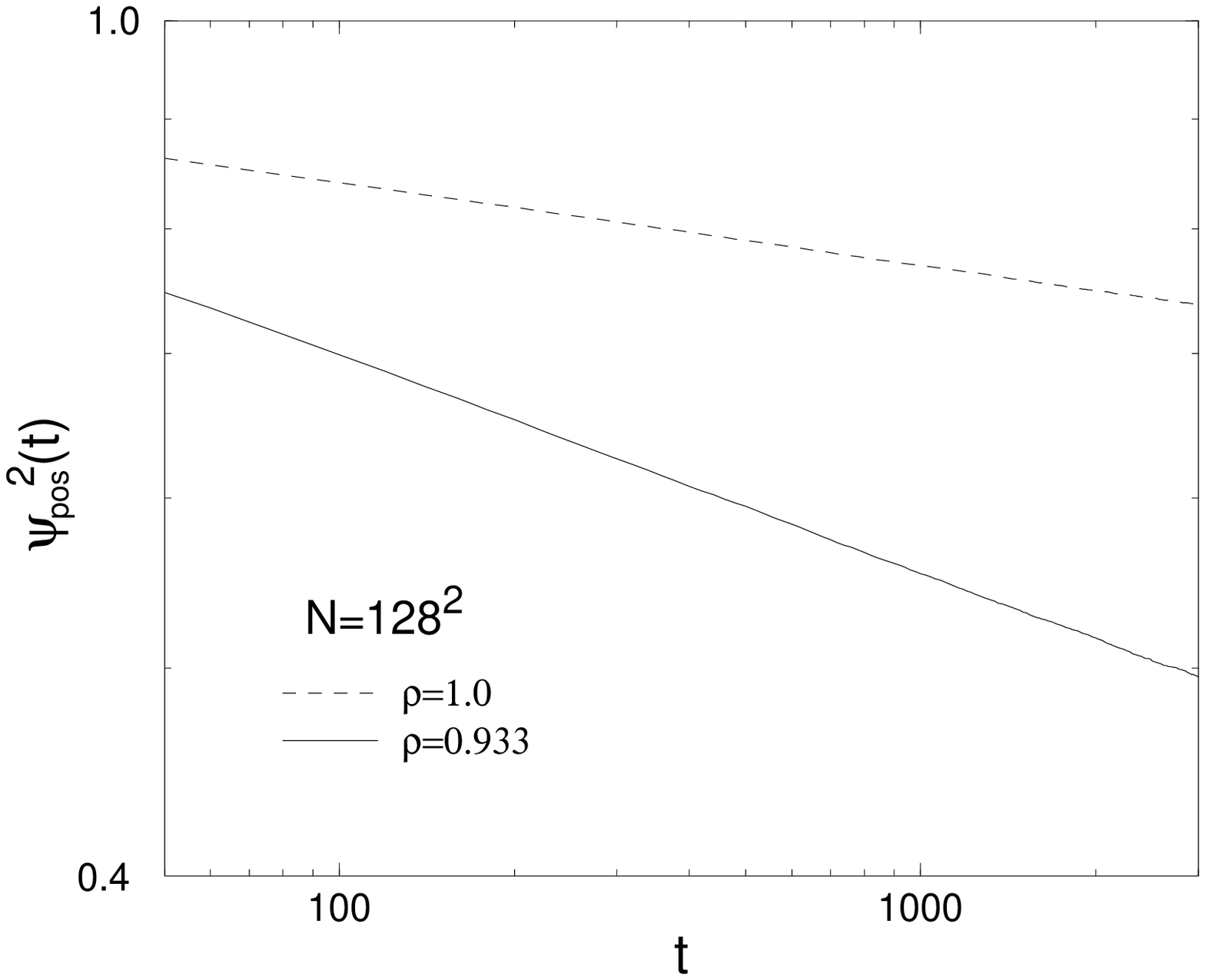}}
\end{center}
\caption{\label{fig_pos_O2}
(a) Time evolution of the second moment of the positional order
parameter ${\psi_{\mathrm{pos}}}^2$ starting from the  ordered state at
$\rho=0.933$. 
%From above, the lines corresponds to $N=8^2$,
%$16^2$, $32^2$ and $64^2$. 
(b)   ${\psi_{\mathrm{pos}}}^2$  
as a function of time for $\rho=0.933$ and $\rho=1.0$.}
\end{figure}
%%%%%%%%%%%%%%%%%%%%%%%%%%%%%%%%%%%%%%%%%%%%%%%%%%%%%%%%%%%%%%%%%%%%%%%% 

In Fig.~\ref{fig_pos_O2} (a) we plot the time evolution of 
${\psi_{\mathrm{pos}}}^2$ at the melting density for different system sizes in
a double logarithmic scale. 
The figure shows that the power law behaviour starts after
a microscopic time scale $t_{\mathrm{mic}}$ of approximately 80 MC
sweeps. For times up to 500  the difference between the systems with
$64^2$ and $128^2$ hard disks is negligible. Therefore, we omitted the data of
$N=128^2$ in the plot. Figure \ref{fig_pos_O2} (b) shows
the time dependence of ${\psi_{\mathrm{pos}}}^2$ for  $N=128^2$
at both densities. In the time interval shown, the slope is
nearly independent of time and yields $\eta/z=0.0990(5)$ at 
$\rho=0.933$ and $\eta/z=0.0385(7)$ at $\rho=1.0$, respectively.

\begin{figure}[b]
\begin{center}
\mbox{\epsfxsize=9.0cm
\epsfbox{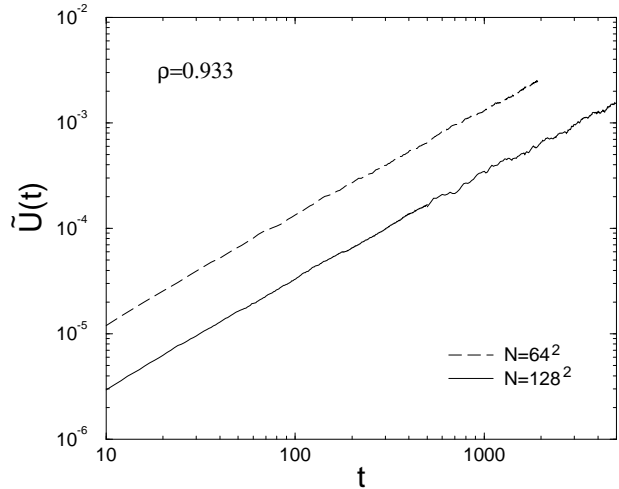}}
\end{center}
\caption{\label{fig_pos_Ut}
Time dependent cumulant $\tilde{U}(t)$ at $\rho=0.933$ for $N=64^2$
and $N=128^2$.}
\end{figure}

To determine $z$ independently, we also measure the time evolution of the
cumulant  $\tilde{U}$. In Fig.~\ref{fig_pos_Ut} we show $\tilde{U}(t)$ 
at the melting density for
two different system sizes. From the slope we get $z=2.01(2)$. The analysis
for $\rho=1.0$ yields $z=2.06(4)$. Thus we get $\eta=0.199(3)$ at
$\rho=0.933$ and $\eta=0.0794(29)$ at $\rho=1.0$. These values coincide with
the results from FSS within statistical errors, as can be seen from Table
\ref{table1}. Thus the data for $\eta$ are confirmed and new
results for $z$ are produced.
%%%%%%%%%%%%%%%%%%%%%%%%%%%%%%%%%%%%%%%%%%%%%%%%%%%%%%%%%%%%%%%%%%%%%%%% 
%%%%%%%%%%%%%%%%%%%%%%%%%%%%%%%%%%%%%%%%%%%%%%%%%%%%%%%%%%%%%%%%%%%%%%%% 
%%%%%%%%%%%%%%%%%%%%%%%%%%%%%%%%%%%%%%%%%%%%%%%%%%%%%%%%%%%%%%%%%%%%%%%%
\begin{table}
\begin{tabular}{c*{3}{r@{.}l}}
\hline
\hline
\multicolumn{1}{c}{ }  & \multicolumn{4}{c}{short-time dynamics} &
\multicolumn{2}{c}{FSS} \\ 
$\rho$  & \multicolumn{2}{c}{$z$} &  
\multicolumn{2}{c}{$\eta$}  &  
\multicolumn{2}{c}{$\eta$} \\
\hline
0.933 &  2&01(2)  &   0&199(3)    &   0&200(2) \\
1.0   &  2&06(4)  &   0&0794(29)  &   0&0791(6) \\
\hline
\hline
\end{tabular}
\caption{ \label{table1}
The critical exponents $z$ and $\eta$ determined from the 
short-time dynamics of the system and the value of $\eta$
measured with FSS methods.}
\end{table}
%%%%%%%%%%%%%%%%%%%%%%%%%%%%%%%%%%%%%%%%%%%%%%%%%%%%%%%%%%%%%%%%%%%%%%%%

In summary, we have performed short-time and equilibrium MC simulations 
of the hard
disk model. The melting density $\rho_{\mathrm{m}}$ and the critical
exponent $\eta$ are determined from simulations in equilibrium. The results
rule out the KTHNY scenario as well as 
a single first-order transition from the solid to the liquid 
phase. The short-time behaviour was used to extract the 
critical exponents $\eta$ and $z$ at $\rho=\rho_{\mathrm{m}}$ and
in the solid phase. The values of $\eta$ coincide 
with those from conventional FSS.
Our simulations have shown that  dynamic relaxation can also be used for
this non-lattice model to measure the critical exponents. 

%%%%%%%%%%%%%%%%%%%%%%%%%%%%%%%%%%%%%%%%%%%%%%%%%%%%%%%%%%%%%%%%%%%%%%%%        
Critical comments by Lothar Sch\"ulke  
are gratefully acknowledged. Especially we benefited from discussion with
Elfie Prinz. This work was supported in part
by the Deutsche Forschungsgemeinschaft under Grant No.\ DFG Schu 95/9-1.

\end{document}